\documentclass[10pt]{article}

\usepackage{graphics}
\usepackage{graphicx}

\usepackage[a4paper, left=35mm,right=35mm,top=34mm,bottom=34mm]{geometry}
\usepackage[utf8]{inputenc}
\usepackage[T1]{fontenc}
\usepackage[english]{babel}

\usepackage{enumerate}
\usepackage{graphicx}
\usepackage{hyperref}
\hypersetup{
    colorlinks=true,
    linkcolor=blue,
    filecolor=magenta,      
    urlcolor=cyan,
}

\usepackage{mathtools,amsthm,amssymb,amsfonts}
\usepackage{algorithm}
\usepackage{algorithmic}
\makeatother
\theoremstyle{plain}

\theoremstyle{definition}

\theoremstyle{remark}

\newtheorem{alg}{Algorithm}

\usepackage{caption} 
\usepackage{listings}
\usepackage{color}
\definecolor{dkgreen}{rgb}{0,0.6,0}
\definecolor{gray}{rgb}{0.5,0.5,0.5}
\definecolor{mauve}{rgb}{0.58,0,0.82}
\usepackage{fancyhdr}

\author{
  {\normalsize Qinmeng Zou}\thanks{CentraleSup\'elec, Universit\'e Paris-Saclay, France.}
  \and
  {\normalsize Guillaume Gbikpi-Benissan}\thanks{CentraleSup\'elec, Universit\'e Paris-Saclay, France.}
	\and
  {\normalsize Fr\'ed\'eric Magoul\`es}\thanks{CentraleSup\'elec, Universit\'e Paris-Saclay, France
    (correspondence, frederic.magoules@hotmail.com).}
		}
\title{Asynchronous Communications Library for the Parallel-in-Time Solution of Black-Scholes Equation}
\date{}

\begin{document}
\maketitle
\thispagestyle{fancy}

\begin{abstract}
\noindent The advent of asynchronous iterative scheme gives high efficiency to numerical computations.
However, it is generally difficult to handle the problems of resource management and convergence detection.
This paper uses JACK2, an asynchronous communication kernel library for iterative algorithms, to implement both classical and asynchronous parareal algorithms, especially the latter.
We illustrate the measures whereby one can tackle the problems above elegantly for the time-dependent case.
Finally, experiments are presented to prove the availability and efficiency of such application.
\end{abstract}


\section{Introduction}

The parareal algorithm is a time domain decomposition method aiming at solving PDEs over a splitting of the original domain into multiple subdomains, over which the equation could be solved independently.
The algorithm is therefore suited for parallel programmin by distributing the computational cost over several processors.
The parallelization with respect to the time variable was earlier attempted by Nievergelt \cite{Nievergelt1964} and by Miranker and Liniger \cite{Miranker1967} for the ordinary differential equations. The parareal algorithm is another extension of decomposition methods in the time direction, first introduced in \cite{Lions2001} that has grown in popularity as it has been successfully applied to a wide variety of problems.
The algorithm consists of two solvers with different time steps, the coarse and the fine, to produce a convergent iterative method for parallel computations.
The coarse solver solves the equation sequentially on the coarse time step while the fine solver uses the results from the coarse solution to solve, in parallel, over the fine time steps.
The solution can therefore be obtained from independent time subdomain, over which the algorithm will iteratively use a coarse and a fine solver to compute the solution on the entire domain.
Several extensions have been proposed following the theoretical analysis.
Bal and Maday in \cite{Bal2002} slightly modified the algorithm, and proposed an analysis of the convergence properties of the algorithm for parabolic equations.

On the other hand, the asynchronous iterative scheme appeared on the horizon since Chazan and Miranker proposed their pioneering ideas applied to the linear systems \cite{Chazan1969}, leading to a prosperity to the high performance applications, such as sub-structuring methods \cite{Magoules2016a} and optimized Schwarz methods \cite{Magoules2017a}.
We are interested in the asynchronous iterations of the parareal method.
Compared to the classical parareal algorithm, the asynchronous version gives further attractive properties including high efficiency and flexibility.
Nevertheless, the implementation of the asynchronous iterative methods involves huge amount of work besides the theoretical analysis, which leads to a frustrating down time for the programming activities.
Fortunately, some libraries have been developed to fill this gap.
JACE \cite{Bahi2004} is a multi-threaded library aiming at the execution of asynchronous iterative algorithms based on Java, followed by a centralized volatility tolerant extension named JaceV \cite{Bahi2007b}.
Recently, they have developed some P2P and decentralized versions to enlarge the applicable scope \cite{Bahi2006, Charr2009}.
Additionally, CRAC \cite{Couturier2007} is another library designed to build the asynchronous applications based on C++.
Finally, JACK \cite{Magoules2017b} is an MPI-based C++ library, which provides many advanced properties designed for parallel iterative algorithms.
A new version called JACK2 has been released.
It gives an improved user-friendly API and several new convergence detection methods with global residual computation.

In this paper we concentrate on the implementation of asynchronous parareal solver based on JACK2 library.
Section \ref{sec:2} recalls the mathematical and computational framework of the parareal algorithm.
In Section \ref{sec:3}, we presents the implementation details of such methods by JACK2.
In Section \ref{sec:4}, we choose the option pricing models to illustrate the experimental results of our asynchronous solver.
Finally, our conclusions and remarks follow in Section \ref{sec:5}.

\section{Mathematical and Computational Framework}
\label{sec:2}

\begin{figure}[!t]
\centering
\includegraphics[width=3.25in]{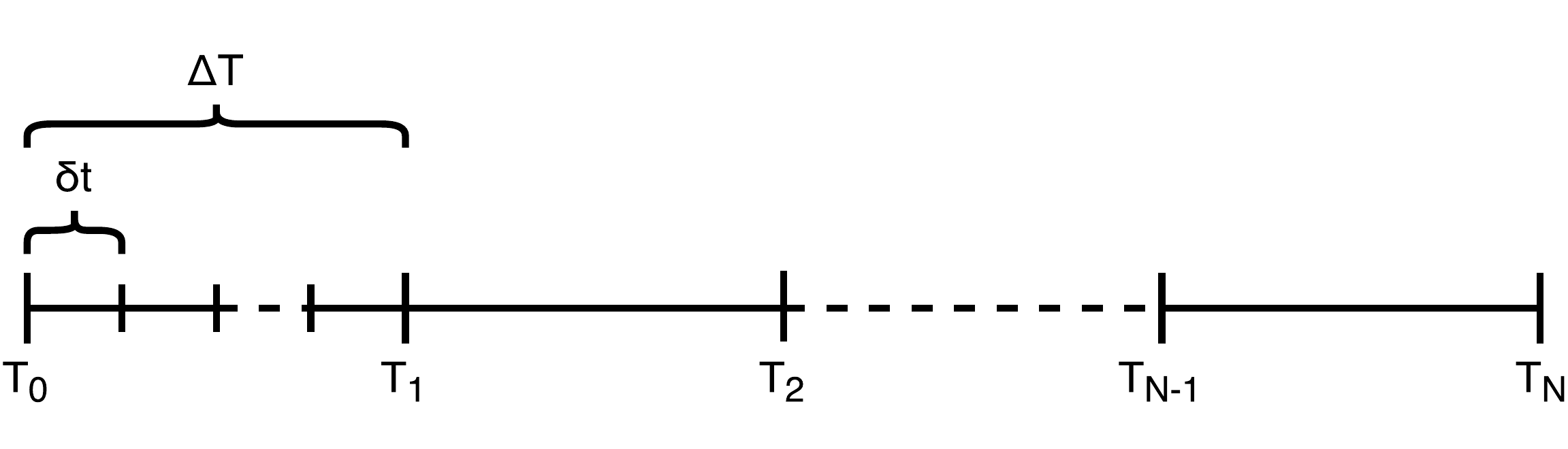}
\caption{Time domain of the parareal scheme}
\label{fig:1}
\end{figure}

Consider a time-dependent problem
\[
\begin{cases}
\frac{\partial u}{\partial t} + \mathcal{L}u = f, & t \in [0, T], \\
u = u_0, & t = 0,
\end{cases}
\]
where $\mathcal{L}$ is a second-order linear elliptic operator.
After a decomposition of the time domain, we arrive at
\begin{equation}
\label{eq:pde}
\begin{cases}
\frac{\partial u_n}{\partial t} + \mathcal{L}u_n = f_n, & t \in [T_n, T_{n+1}], \\
u_n = \lambda_n, & t = T_n,
\end{cases}
\end{equation}
where the time domain is defined in a series of $[T_n, T_{n+1}]$, with $T_n = n\Delta T$, $n = 0, \dots, N$.
It is easily seen that the problem (\ref{eq:pde}) is also an exact model.
Nevertheless, we often choose an accurate enough sub-interval $\delta t$ to approximate the original problem as shown in Figure \ref{fig:1}.
We note here that the limit of each subproblem satisfies
\[
\lambda_{n+1} = \lim_{\epsilon\rightarrow 0}u_n(T_{n+1} - \epsilon),
\]
which enables us to execute a continuous computation.
A well known model proposed in \cite{Lions2001} can be described as the following
\[
\lambda_{n+1}^{k+1} = G(\lambda_n^{k+1}) + F(\lambda_n^k) - G(\lambda_n^k),
\]
where $G$ is a coarse propagator, and $F$ is a fine propagator.
Accordingly, $\Delta T$ represents a coarse time step, whereas $\delta t$ is a fine time step.
We notice that this is a synchronous parareal iterative scheme, over which one can come up with the algorithmic model forthwith
\begin{alg}
\label{alg:cpr}
(Classical Parareal Algorithm)

\begin{algorithmic}
\STATE $n\coloneqq$ rank of current processor
\STATE $\lambda_0 = u_0$
\FOR{$i=0$ \TO $n-1$}
\STATE $\lambda_0 = G(\lambda_0)$
\ENDFOR
\STATE{$w=G(\lambda_0)$}
\WHILE{not convergence}
\STATE{$v = F(\lambda_0)$}
\STATE{wait for the update of $\lambda_0$ from processor $n-1$}
\STATE{$\tilde{w} = G(\lambda_0)$}
\STATE{$\lambda = \tilde{w} + v - w$}
\STATE{send $\lambda$ to processor $n+1$ as $\lambda_0$}
\STATE{$w = \tilde{w}$}
\ENDWHILE
\end{algorithmic}
\end{alg}
\noindent Evidently, the first processor has only successor, whereas the last one has none but predecessor.

We now turn to the asynchronous mode.
Formally, we view parareal as a two-stage problem which can be formalized carefully using the asynchronous scheme, illustrated as below
\[
\lambda_{n+1}^{k+1} =
\begin{cases}
G(\lambda_n^{\mu_n(k)}) + F(\lambda_n^{\rho_n(k)})
- G(\lambda_n^{\rho_n(k)}), & n \in P^k, \\
\lambda_{n+1}^k, & n \notin P^k,
\end{cases}
\]
where $P^k \subseteq \{1,\dots,p\}$ and $P^k \neq \emptyset$, with the conditions $0 \le \mu_j^i(k) \le k+1$, $0 \le \rho_j^i(k) \le k$.
We assume that each processor keeps updating rather than resting permanently.
Furthermore, the elements used by this processor will be renewed from time to time.
Similarly, such computational model can be written as
\begin{alg}
\label{alg:apr}
(Asynchronous Parareal Algorithm)

\begin{algorithmic}
\STATE $n\coloneqq$ rank of current processor
\STATE $\lambda_0 = u_0$
\FOR{$i=0$ \TO $n-1$}
\STATE $\lambda_0 = G(\lambda_0)$
\ENDFOR
\STATE{$w=G(\lambda_0)$}
\WHILE{not convergence}
\STATE{$v = F(\lambda_0)$}
\IF{detect $\lambda_0$ from processor $n-1$}
\STATE{update $\lambda_0$}
\ENDIF
\STATE{$\tilde{w} = G(\lambda_0)$}
\STATE{$\lambda = \tilde{w} + v - w$}
\STATE{send $\lambda$ to processor $n+1$ as $\lambda_0$}
\STATE{$w = \tilde{w}$}
\ENDWHILE
\end{algorithmic}
\end{alg}
\noindent It is seen that Algorithm \ref{alg:apr} has no waiting process in the internal loop, whereas Algorithm \ref{alg:cpr} requires a latency time, which might cause a waste of time on coordination, but gain through the number of iterations.

\section{Implementations}
\label{sec:3}

\subsection{Preprocessing}

The parareal algorithm is indeed a special case of domain decomposition methods, since each time frame only depends upon its predecessor, and essentially needed by its successor.
We separate the neighbors into the outgoing links and the incoming links, and thus write the code as Listing \ref{lst:cg}.
It is seen that \textit{sneighb\_rank} and \textit{rneighb\_rank} have respectively one element in the communication graph.
Take note that the first processor has no predecessor, therefore $\textit{numb\_rneighb}=0$; while the last one has no successor, obviously, with $\textit{numb\_sneighb}=0$.
\begin{lstlisting}[caption=Communication graph, label=lst:cg]
/* template <typename T, typename U> */
// T: float, double, ...
// U: int, long, ...
U numb_sneighb = 1;
U numb_rneighb = 1;
U* sneighb_rank = new U[1];
U* rneighb_rank = new U[1];
\end{lstlisting}

The key idea behind this project is that there exist two levels over the computation process.
The upper object is the instance of \textit{Parareal}, which provides the interface of JACK2's power, whereas the lower level is the \textit{PDESolver}, used to solve the equations.
We notice that it is easy to enlarge the number of solvers without changing the code of \textit{Parareal}.
Accordingly, we need two instances of \textit{PDESolver} in \textit{Parareal} to obtain respectively the coarse results and the fine results, illustrated in Listing \ref{lst:ps}.
\begin{lstlisting}[caption=PDE solver in Parareal, label=lst:ps]
/* template <typename T, typename U> */
PDESolver<T,U> coarse_pde;
PDESolver<T,U> fine_pde;
Vector<T,U> coarse_vec_U;
Vector<T,U> fine_vec_U;
Vector<T,U> vec_U; // solution vector.
Vector<T,U> vec_U0; // initial vector.
\end{lstlisting}
\textit{coarse\_vec\_U} is the solution of \textit{coarse\_pde}, while \textit{fine\_vec\_U} is the solution of \textit{fine\_pde}.
\textit{vec\_U} is the final solution, which is also the output of \textit{Parareal} solver.
Finally, \textit{vec\_U0} is the incoming value for both \textit{coarse\_pde} and \textit{fine\_pde}.

\subsection{Overview of the Iterations}

We now present the iterative process of parareal algorithm.
Notice that the beginning parts of Algorithm \ref{alg:cpr} and Algorithm \ref{alg:apr} are the same, we can therefore write in a unified scheme in Listing \ref{lst:initp}.
\begin{lstlisting}[caption=Initialization of parareal iterative process, label=lst:initp]
/* template <typename T, typename U> */
for (U i = 0; i < rank; i++) {
	coarse_pde.Integrate();
	vec_U0 = coarse_vec_U;
}
coarse_pde.Integrate();
vec_U = coarse_vec_U;
\end{lstlisting}
Let us mention here that Listing \ref{lst:initp} gives us the exact position of \textit{vec\_U0} and the initial estimation of $G(\lambda_n^0)$.
Nevertheless, we need to distinguish the synchronous and the asynchronous parareal algorithm afterwards.

\subsubsection{Synchronous mode}

We follow the scheme of Algorithm \ref{alg:cpr} except for the convergence detection mode, since there is a rule of thumb to lighten the communication.
It is easily seen that the processor $n$ will stop updating after $(n+1)^{\text{th}}$ iteration.
Hence, we can apply a supplementary condition to the judging area in the synchronous mode, with the final code shown as Listing \ref{lst:sip}.
\begin{lstlisting}[caption=Synchronous iterative process, label=lst:sip]
// -- synchronous parareal iterations
res_norm = res_thresh;
numb_iter = 0;
while (res_norm >= res_thresh &&
				  numb_iter < m_rank) {
	fine_pde.Integrate();
	comm.Recv();
	coarse_vec_U_prev = coarse_vec_U;
	coarse_pde.Integrate();
	vec_U_prev = vec_U;
	vec_U = coarse_vec_U + fine_vec_U - coarse_vec_U_prev;
	comm.Send();
	// -- |Un+1<k+1> - Un+1<k>|
	vec_local_res = vec_U - vec_U_prev;
	(*res_vec_buf) = vec_local_res.NormL2();
	comm.UpdateResidual();
	numb_iter++;
}
\end{lstlisting}
We note that it is necessary to copy the values of \textit{recv\_buf} to \textit{vec\_U0} and the values of \textit{vec\_U} to \textit{send\_buf} whenever needed.
However, we can also ease the process by setting directly the two vectors as communication buffers.
In practice, the choice relates to the mathematical library used throughout the project.
Moreover, using the results of Listing \ref{lst:sip}, it is also shown in Listing \ref{lst:sfp} that we need to finalize the fine solution and wait for the global termination.
\begin{lstlisting}[caption=Synchronous finalized process, label=lst:sfp]
// -- computes sequential fine solution
if (res_norm >= res_thresh) {
	fine_pde.Integrate();
	vec_U = fine_vec_U;
	comm.Send();
	// -- wait for global termination
	(*res_vec_buf) = 0.0;
	while (res_vec_norm >= res_thresh) {
		comm.UpdateResidual();
		numb_iter++;
	}
}
\end{lstlisting}

\subsubsection{Asynchronous mode}

The asynchronous parareal scheme can be implemented similarly with Listing \ref{lst:sip} except for the matching conditions.
There is no more facile code in practice, and we present the process as Listing \ref{lst:aip}.
From the entry of \textit{lconv\_flag}, JACK2 can interact with the asynchronous process and give back results as \textit{res\_norm}.
\begin{lstlisting}[caption=Asynchronous iterative process, label=lst:aip]
// -- asynchronous parareal iterations
res_norm = res_thresh;
numb_iter = 0;
while (res_norm >= res_thresh) {
	fine_pde.Integrate();
	comm.Recv();
	coarse_vec_U_prev = coarse_vec_U;
	coarse_pde.Integrate();
	vec_U_prev = vec_U;
	vec_U = coarse_vec_U + fine_vec_U - coarse_vec_U_prev;
	comm.Send();
	// -- |Un+1<k+1> - Un+1<k>|
	vec_local_res = vec_U - vec_U_prev;
	(*res_vec_buf) = vec_local_res.NormL2();
	lconv_flag = ((*res_vec_buf) < res_thresh);
	comm.UpdateResidual();
	numb_iter++;
}
\end{lstlisting}

\section{Experimental Results}
\label{sec:4}

In this section, we give the experiments for the asynchronous parareal scheme, based on the Black-scholes equation, which is a well-known time-dependent equation in the domain of option pricing. Consider the following problem
\[
\frac{\partial V}{\partial t} +  rS\frac{\partial V}{\partial S} +
\frac{1}{2}\sigma^2S^2\frac{\partial^2 V}{\partial S^2} = rV,
\]
where $V$ is the option price, depending on stock price $S$ and time $t$.
Volatility $\sigma$ and risk-free interest rate $r$ are the constant parameters.
Then, we employ some trivial variable substitutions to obtain the heat equation
\begin{equation}
\label{eq:heat}
\frac{\partial u}{\partial \tau} = \frac{\partial^2 u}{\partial x^2},\quad
x\in\mathbb{R},\ \tau\in[0, \frac{T\sigma^2}{2}],
\end{equation}
where $T$ is the time to maturity, with initial and boundary conditions
\[
\begin{cases}
u(x,0) = \max(e^{\beta x}-e^{\alpha x}, 0),
& x\in\mathbb{R}, \\
u(x,\tau) \sim 0\ as\ x \rightarrow \pm\infty, & \tau\in[0, \frac{T\sigma^2}{2}].
\end{cases}
\]
where $\alpha = \frac{1}{2}(\kappa-1)$, $\beta = \frac{1}{2}(\kappa+1)$, $\kappa = \frac{2r}{\sigma^2}$.
Obviously, equation (\ref{eq:heat}) can be applied to the classical and the asynchronous parareal scheme.
We note here that the input variables are the current stock price $S$ and the exercise price $E$.
Finally, we broaden the spatial domain and let the dimension correspond to the precision of our problem.

The mathematical operations are supported by Alinea library \cite{Magoules2015a}, which is implemented in C++ for both central processing unit and graphic processing unit devices. 
It includes several linear algebra operations \cite{Ahamed2016c} and numerical linear algebra solvers \cite{Magoules2015b}, \cite{Ahamed2016d}, \cite{Magoules2015c}.

The experiments are exercised on a SGI ICE X cluster connected with InfiniBand.
Each node consists of two Intel Xeon E5-2670 v3 2.30 GHz CPUs with SGI-MPI 2.14 installed.
We assume that $\sigma=0.2$, $r=0.03$, $\delta t=0.001$ with 250 sub-intervals.
Given $S=100$, $E=80$, $N=16$, several average results are reported in Table \ref{tab:res} with Approximate Option Prices $V_a$, Exact Option Prices $V_e$, Absolute Error $\epsilon_a$ and Relative Error $\epsilon_r$.
\begin{table}[!ht]
\caption{Results of Asynchronous Parareal Scheme
($\sigma=0.2$, $r=0.03$, $\delta t=0.001$, $S=100$, $E=80$, $m=250$, $N=16$)}
\label{tab:res}
\centering
\begin{tabular}{|c||c|c|c|c|c|}
\hline
$\Delta T$ & $V_a$ & $V_e$ & $\epsilon_a$ & $\epsilon_r$ & Time \\
\hline
0.05 & 23.9476 & 23.9426 & 0.0050 & 0.0002 & 0.547 \\
0.15 & 31.5512 & 31.5477 & 0.0035 & 0.0001 & 1.663 \\
0.25 & 37.7225 & 37.7192 & 0.0033 & 0.0001 & 2.881 \\
0.35 & 42.9995 & 42.9960 & 0.0035 & 0.0001 & 3.830 \\
0.45 & 47.6375 & 47.6339 & 0.0036 & 0.0001 & 4.814 \\
\hline
\end{tabular}
\end{table}
It is seen that the asynchronous parareal algorithm implemented by JACK2 is efficient and accurate.

\section{Conclusions}
\label{sec:5}

In this paper, we illustrated the implementation of synchronous and asynchronous parareal algorithm using JACK2, an asynchronous communication kernel library for iterative algorithms.
We discussed the particularity of such time-dependent problem, which leads to a rather different configuration than the general parallel context.
As expected, the experimental results at the end verified the correctness of the asynchronous scheme.

\bibliography{ref}
\bibliographystyle{abbrv}

\end{document}